\documentclass[12pt]{article}
\usepackage{amsmath}
\usepackage{amssymb}
\usepackage{amsfonts}
\usepackage{pstricks}
\usepackage{graphicx}
\usepackage[small,bf]{caption}

\newcommand{\mgut}{\ensuremath{M_{GUT}}}
\newcommand{\mc}{\ensuremath{M_{C}}}
\newcommand{\ms}{\ensuremath{M_{S}}}
\newcommand{\mext}{\ensuremath{M_{EX3}}}
\newcommand{\mexd}{\ensuremath{M_{EX2}}}
\newcommand{\eps}{\ensuremath{\epsilon_{3}}}
\newcommand{\mgrav}{\ensuremath{m_{3/2}}}

\def\beq{\begin{equation}}
\def\eeq{\end{equation}}
\def\beqn{\begin{eqnarray}}
\def\eeqn{\end{eqnarray}}

\begin{document}

\begin{flushright}
OHSTPY-HEP-T-10-007 \\
\end{flushright}
\vskip 2cm

\begin{center}
{\large \bf Gauge Coupling Unification in Heterotic String Models
with Gauge Mediated
SUSY Breaking\\[2ex]}
\vspace*{5mm} \vspace*{1cm}
\end{center}
\vspace*{5mm} \noindent \vskip 0.5cm \centerline{ Archana
Anandakrishnan and Stuart Raby} \vskip 1cm \centerline{ \em
Department of Physics, The Ohio State University,} \centerline{\em
191 W.~Woodruff Ave, Columbus, OH 43210, USA} \vskip2cm

\centerline{\bf Abstract}
\vskip .3cm
We calculate the weak scale MSSM spectrum starting from a heterotic string theory compactified on an
anisotropic orbifold. Supersymmetry breaking is mediated by vector-like exotics that arise naturally in 
heterotic string theories. The messengers that mediate SUSY breaking come in incomplete GUT multiplets and give rise
to non-universal gaugino masses at the GUT scale. Models with non-universal gaugino masses at the GUT scale have the 
attractive feature of allowing for precision gauge coupling unification at the GUT scale with negligible contributions
from threshold corrections near the unification scale. The unique features of the MSSM spectrum are
light gluinos and also large mass differences between the lightest and the next-to-lightest neutralinos and charginos which could lead
to interesting signatures at the colliders.
\vskip .3cm

\newpage

\section{Introduction}
Grand unification of the fundamental forces is a very appealing
idea. It was noticed in as early as 1974, that when the couplings of
the three fundamental interactions are run to high energies, they
seem to meet at a point \cite{Georgi:1974sy},\cite{Georgi:1974my}. Supersymmetry is required for precise
unification \cite{Dimopoulos:1981yj}, without which there is a discrepancy of about 12
$\sigma$ \cite{Amsler:2008zzb}. With supersymmetry, assuming
universal scalar and gaugino masses at the GUT scale, precision
electroweak data requires \cite{kazakov}, for the strong coupling constant to match
experiments, that $\alpha_{3}$ be about $3 - 4 \%$ smaller than
$\alpha_{1}$ and $\alpha_{2}$ at the GUT scale. This conflict
between the coupling constants at the GUT scale can be eased by
including `threshold corrections' from extra states around the GUT
scale.

In grand unified theories, the Higgs fields have to respect the GUT
symmetry and thus existence of Higgs doublets also implies the
existence of Higgs triplets.  In order to avoid rapid proton decay
the triplets necessarily have mass greater than the GUT scale which
introduces some unpleasantness into these SUSY GUT theories. In
addition, complicated symmetry breaking potentials are required to
break the GUT symmetry. Theories with extra dimensions have gained
popularity in this respect, since they can eliminate some of the
problems with 4D SUSY GUTs.  In theories with extra dimensions,
connection is made to the low-energy world, by compactifying the
extra dimensions. The choice of boundary conditions then can lead to
natural and simple solutions to the problems hindering SUSY GUTs.
The threshold corrections required to match precision electroweak
data can come from massive states around the GUT scale and from
Kaluza-Klein states living between the compactification scale of the
extra dimensions, \mc\ and the cut-off scale, $M_{*}$; in string
theory $M_*$ is the string scale, $M_S$.

Recent searches for the MSSM from heterotic string theory have
yielded interesting results. Orbifold compactifications of the
$E_{8} \times E_{8}$ heterotic string theory have been shown to
yield realistic models that include the gauge group and the matter
content of the MSSM \cite{gauge-higgs, minilandscape}. In addition,
the models also have vector-like exotics with Standard Model charges
that obtain mass in the supersymmetric limit. They may couple to the
SUSY breaking field, and mediate supersymmetry breaking. The
mechanism of supersymmetry breaking plays a very important role in
understanding the low-energy spectrum, and in this case the
possibility of \textit{Gauge Mediated Supersymmetry breaking} (GMSB)
\cite{Giudice:1998bp}. In GMSB, the gauginos receive mass at
one-loop as: \beq
 M_{i} \sim \frac{\alpha_{i}}{4 \pi} \frac{\langle F \rangle}{\langle
 M_\phi
\rangle} \eeq where, $\phi$ is the messenger field with mass,
$M_\phi$, and $\langle F \rangle$ is the SUSY breaking VEV. Thus, a
heavier messenger (in this case, the exotics) corresponds to a
lighter gaugino.

It was shown earlier in \cite{Dundee:2008ts} that light(of the order
$10^{9} - 10^{13}$ GeV) vector-like exotic states were required for
gauge coupling unification, assuming the standard scenario with
universal gaugino masses at the GUT scale and threshold corrections
of about -3\%. Solutions to gauge coupling unification were
constrained by the bounds on proton decay. It was also assumed that
all the vector-like exotics obtain mass at the same scale. The
exotics come in incomplete GUT multiplets and hence, in general
could obtain mass at different scales. In this work, we generalize
the solutions, allowing the exotics that carry SU(3) and SU(2)
charges to obtain mass at different scales.  We build a
\textit{consistent} MSSM spectrum at the weak scale with these
exotic messengers.  We find that this generalization increases the
number of solutions satisfying gauge coupling unification. In
addition, the weak scale spectrum now allows unification with
moderate or even zero threshold corrections at the GUT scale. The
low energy spectrum in such a case has light gluinos which should be
detected at the Tevatron and/or LHC.

\section{Gauge Mediated Supersymmetry Breaking}
Gauge Mediated SUSY Breaking(GMSB) \cite{gaugemediation} models have
chiral supermultiplets called \textit{messenger} fields that mediate
supersymmetry breaking. The messenger fields carry SU(3) $\times$
SU(2) $\times$ U(1) charges and hence couple to the matter fields of
the MSSM through the usual SU(3) $\times$ SU(2) $\times$ U(1) gauge
interactions. The messenger fields are very massive at some scale,
denoted by $M_{mess}$. Sources of flavor violation near the messenger
scale are given by (4+d)-dimension operators that are suppressed by
$\frac{1}{(M_{mess})^{d}}$. Hence, the major source of flavor violation
is due to Yukawa couplings, similar to the Standard Model(SM). This
suppression of flavor changing neutral currents (FCNC) is the most
attractive feature of GMSB.

In string theories where the extra-dimensions are compactified on an
orbifold, there exist extra vector-like non-standard model
particles, usually called ``exotics''. These exotics need to be heavy in
order for them to decouple from the low-energy theory. The exotics carry
charges under the SM gauge group, and hence are perfect candidates
to mediate supersymmetry breaking via gauge mediation. We build a
\textit{consistent} MSSM spectrum at the weak scale with these
exotic messengers. By consistent, we mean that if we start at the
highest scale in the model and run the coupling constants and soft
SUSY breaking parameters all the way down to the weak scale,
integrating out heavy states during this running, we must end up
with the coupling constants that match the experimental values at
the weak scale. Since this requires knowledge of the spectrum of
exotics and Kaluza-Klein modes, as well as the
MSSM spectrum, we perform our analysis in two steps:\\[5pt]

\textbf{Step \#1} We concentrate on a class of models based on SU(6)
gauge-Higgs unification in 5D \cite{gauge-higgs, minilandscape}.
Starting from heterotic string theory compactified on an anisotropic
orbifold, $T^{6} / \mathbb{Z}_{6}$-II and applying the 
`Phenomenological Priors' - Inequivalent models with the SM gauge
group, 3 SM families, Higges, and non-anomalous $U(1)_{Y} \subset
SU(5)$; the authors end up with 15 models consisting of low energy spectrum
that is similar to that of MSSM. In addition, the spectrum consists
of heavy vector-like exotics that decouple from the low-energy
theory. Gauge coupling unification was studied \cite{Dundee:2008ts}
in 2 out of the above 15 models \cite{minilandscape} - ``Model 1A''
and ``Model 2''. The matter content of both these models are very
similar and is summarized in Table \ref{tab:all_exotics}. For the
gauge couplings to unify in the heterotic orbifold theory, it was
noted in Ref.\cite{Dundee:2008ts} that there had to be at least
$\vec{n} = (n_{3}, n_{2}, (n_{1}, n'_{1})) $ `light' exotics at some
intermediate scale $M_{EX}$, below the 4D unification(GUT) scale.
This scale, $M_{EX}$ was determined by matching the Renormalization
Group Equations(RGE) from the two theories - heterotic orbifold
model and the 4D MSSM at some low energy scale $\mu$, where both the
theories predict the same running for the couplings. In the 4D MSSM,
the gauge couplings unify at the GUT scale with some threshold
corrections from new physics near the GUT scale whereas, on the
heterotic side, the gauge couplings unify at the string scale, \ms
(See Fig.\ref{running}). The analysis in Ref.\cite{Dundee:2008ts}
was done in the context of a minimal scenario where all the light
exotics obtained mass at the same scale and to accommodate precision
electroweak data, a -3 \% threshold correction was assumed at the
GUT scale.

The exotics come in incomplete SU(5) GUT multiplets and in general
could obtain mass at different scales. We therefore relax the
previous assumption that the light exotics obtain mass at the same
scale and allow those that carry SU(3) and SU(2) charges to obtain
mass at different scales. This leads to non-universal gaugino masses
at the GUT scale, as a consequence of which, the threshold
corrections required to match precision data at this scale is no
longer of order -3\%.  The GUT scale threshold correction is a
priori a free parameter which depends on the spectrum of states near
the GUT scale. However, it's value needs to be fixed by evaluating
the 2-loop RGE running from the string scale to the weak scale and
including one loop threshold corrections at both the weak and the
GUT scales, self-consistently. This is done in the next step. The
new intermediate scales, \mext\ and \mexd\ are determined
self-consistently using the RGEs. The details of the calculation of
the light exotics mass spectrum is given in Appendix \ref{rge}.\\[5pt]

\textbf{Step \#2} Once the exotic masses are determined, the soft
SUSY breaking terms are calculated at this scale, i.e. the messenger
scale \cite{Giudice:1998bp}.   We then use SOFTSUSY
\cite{Allanach:2001kg} to run them down to the weak
scale.\footnote{Note, SOFTSUSY runs from the presumed 4D GUT scale
to the weak scale.  In our case the soft masses are only determined
at the messenger scale. The error made by matching the 5D theory to
the 4D theory at the messenger scale is however, small. Using the
fact that up to 1-loop, the ratio $M_{i}$/$\alpha_{i}$ = constant,
we calculate the soft-masses at the GUT scale where they get small
2-loop corrections. As shown in Appendix. \ref{2lc}, the corrections
to the gaugino masses are found to be less than 1\% and can be
neglected.  A detailed discussion of the soft SUSY breaking masses
is in Section \ref{sec:soft}.} SOFTSUSY uses the 2-loop RG running
to determine the weak scale MSSM spectrum and the 1-loop weak scale
threshold corrections. The MSSM parameters are then run back again
to the GUT scale and the GUT scale threshold corrections are
calculated, i.e. the (output) values fixed by SOFTSUSY (see Eq.
(\ref{4drg})). We compare this value of the GUT threshold
corrections with the (input) value determined independently by the
exotic mass spectrum (see Eq. (\ref{orbrge})). We vary the arbitrary
parameters of the orbifold string theory and save only those cases
where the input (determined by the exotic mass spectrum) and output
(required by the low energy MSSM spectrum) threshold corrections
match.  We also only keep cases consistent with the bound on the
proton lifetime and a lower bound on the Higgs mass.

\subsection{\label{sec:soft} Soft Masses}
The exotics of the orbifold theory mediate supersymmetry breaking by
acting as messengers of gauge mediation. Due to the gauge
interactions of the messengers, soft terms are generated at the
messenger scale.  We assume that the gravity-mediated SUSY breaking
contributions to gaugino masses are much smaller than the
gauge-mediated contribution. There can be anomaly mediated
contributions to the soft masses, proportional to the gravitino
mass, \mgrav, or dilaton contributions proportional to
$F_S/M_{Pl}$.\footnote{We assume the SUSY breaking dilaton VEV,
$F_S$, is negligibly small. Kahler and complex structure moduli
(denoted generically by $T$) contribute to gaugino masses via one
loop corrections to the gauge kinetic function with scale set by
$F_T/M_{Pl}$. $F$ is then assumed to be a linear combination of
geometric moduli and chiral matter moduli.}  We allow for a large
gravitino mass of the order of a few TeV. However, if the ratio,
$\frac{F}{M_{EX}} \gg \mgrav$, we can ignore the gravitino
contribution to the gaugino masses. At one loop, the gauginos masses
are given by: \beq M_{i} = b^{EX3}_{i} \ \frac{\alpha_{i}}{4 \pi}
\frac{F}{\mext} + b^{EX2}_{i} \ \frac{\alpha_{i}}{4 \pi}
\frac{F}{\mexd} \label{gaugino} \eeq where F is the SUSY breaking
VEV, which at this point is chosen to be arbitrary.

The scalars obtain mass at two-loops, and the dominant contribution
to their mass is from the gravitino. In addition, in string models,
it is natural to have an anomalous $U(1)_{X}$ gauge interaction.
Such interactions can add an additional Fayet-Iliopoulos D-term to
the scalar potential. In such cases, the scalar masses can receive a
contribution from the D-term that is of the same order as the gauge
mediation contribution. This was discussed in \cite{Raby:1998bg}
where the contribution to scalar masses was modeled by a term: \beq
\delta m^{2}_{\phi_{i}} = d \ Q_{i}^{X} M_{2}^{2} \eeq with, $d$, an
arbitrary parameter and, $Q^{X}_{i}$, the $U(1)_{X}$ charge of the
field $\phi_{i}$. For the matter fields this charge is taken to be
+1, and for the Higgs fields, it is set equal to -2; i.e. $U(1)_X$
is the U(1) in $SO(10)$ commuting with $SU(5)$. $M_{2}$ represents
the wino mass calculated earlier in Eq. (\ref{gaugino}). With
contributions from the gravitino, gauge mediation, and the D-term,
the scalar masses are given by: \beqn m^{2}_{\phi_{i}} = \mgrav^{2}
+ 2 \left( b^{EX3}_{3}\ \frac{\alpha_{3}}{4 \pi} \frac{F}{\mext}
\right)^{2} C_{3} (i) + 2\left( b^{EX2}_{2} \ \frac{\alpha_{2}}{4
\pi} \frac{F}{\mexd}  \right)^{2} C_{2} (i) \nonumber \\ + 2\left(
\frac{\alpha_{3}}{4 \pi} \left( b^{EX3}_{1}\ \frac{F}{\mext} +
b^{EX2}_{1}\ \frac{F}{\mexd} \right) \right)^{2} C_{1} (i) + d
Q_{i}^{X} M_{2}^{2} \label{scalar} \eeqn where, $C_{i}$s represent
the quadratic Casimir invariants \cite{Martin:1997ns}. The low
energy spectrum is now computed for different values of $\vec{n}$,
\eps, $F$, \mgrav, and d.  Table \ref{gut_spectrum} shows the GUT
scale parameters for four sample points.

\begin{table}[[ht]
\centering
 \caption{\label{gut_spectrum}The GUT scale parameters for four different
cases. $g_{string}$ is discussed in Appendix \ref{gcoupling}. Dimensionful quantities in
units of GeV unless specified.}
 \vspace{2mm}
\begin{tabular}{|c|c|c|c|c|}
\hline
 \textbf{Observable} & \textbf{Case 1} & \textbf{Case 2} & \textbf{Case 3} & \textbf{Case 4}\\ \hline
 $\vec{n}$ & (4,2,(2,1))& (4,2,(2,1))& (4,2,(1,1))& (4,2,(2,0)) \\
\mgrav& 4 TeV &  10 TeV &  10 TeV &  4 TeV \\
 d & 0 & 5 & 5 & 1\\ \hline
$g_{string}$  & 0.99412 & 0.99604 & 0.8233 & 0.8588 \\ \hline
 \ms & 6.04 $\times 10^{17} $  & 6.05 $\times 10^{17} $ & 7.39 $\times 10^{17} $& 7.27 $\times 10^{17} $ \\
\mc & 1.2 $\times 10^{16} $  & 1.2 $\times 10^{16} $ & 3.2 $\times 10^{16} $& 2.8 $\times 10^{16} $\\
 \mext & 5.03 $\times 10^{13}$ & 1.10 $\times 10^{14}$ & 1.07 $\times 10^{14}$ & 5.05 $\times 10^{13}$ \\
 \mexd & 1.69 $\times 10^{13}$ &  8.54 $\times 10^{13} $ &  5.35 $\times 10^{13} $ &  8.87 $\times 10^{12} $\\ \hline
 \mgut & 2.5 $\times 10^{16}$ &   2.0 $\times 10^{16}$ &  3.25 $\times 10^{16}$ &  1.75 $\times 10^{16}$ \\
 \eps & -2.5 \%& 0 \% & -2.5 \% & -0.5 \% \\
F & $1.0 \times 10^{18} \text{GeV}^{2}$& $1.0 \times 10^{18} \text{GeV}^{2}$ & $1.0 \times 10^{18} \text{GeV}^{2}$ & $1.0 \times 10^{18} \text{GeV}^{2}$\\ \hline
 $M_{3}$(\mgut) & 257.296& 155.269 & 120.882 & 260.894\\
 $M_{2}$(\mgut) & 392.844& 600.865 & 119.793 &747.307 \\
 $M_{1}$(\mgut) & 124.900&128.947 & 39.666 & 260.894 \\
\hline
\end{tabular}
\end{table}

\section{Features of the Spectrum}
\begin{figure}
\centering
  \includegraphics[width=9.5cm]{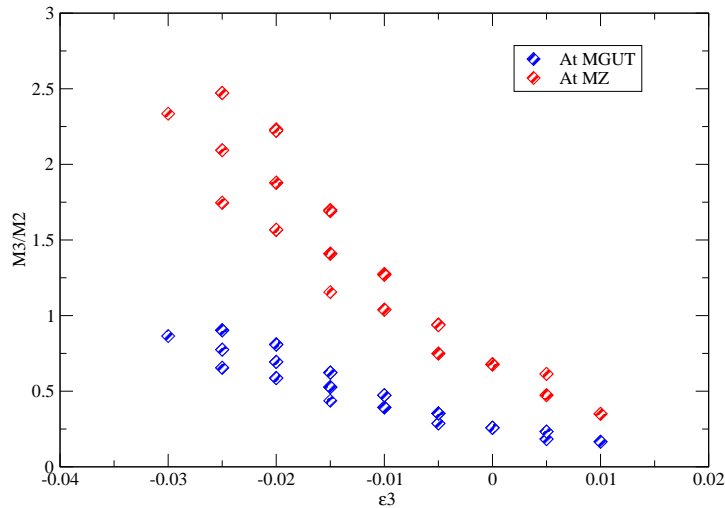}
\caption{\label{m3bym2}The scatter plot of consistent points for the case,
$\vec{n} = (4, 2, (2, 1))$. The gravitino mass, \mgrav\ was taken to be 4 TeV. The running of
the couplings depend only on the ratio $M_{3}/M_{2}$ since the scalars are very heavy. We find
an anti-correlation between this ratio and the value of \eps. Precision unification favors $M_{3}/M_{2} \sim 0.3$ at the
GUT scale.}
\end{figure}

The MSSM spectrum is calculated using SOFTSUSY
\cite{Allanach:2001kg}. For the four cases shown in Table
\ref{gut_spectrum}, the spectrum from SOFTSUSY is shown in Table.
\ref{mssm_spectrum}.  \\ [5pt] \textit{Non-universal gaugino
masses}: The split exotics give rise to non-universal gaugino masses
at the GUT scale, as is clear from Eq.(\ref{gaugino}). As a result
of this non-universality, the GUT scale threshold corrections
required to match the precision electroweak data need not
necessarily be of order -3 \% . In fact, we notice that it is
possible to obtain precision unification when $M_{3} < M_{2}$. This
requirement was observed in \cite{Raby:2009sf} in a variety of SUSY
breaking scenarios including a Higgs-messenger mixing model where
SUSY is broken via gauge mediation. The weak scale MSSM spectrum now
has a light gluino. Case 2 in Table \ref{mssm_spectrum} illustrates
this feature of the gluino being the second lightest sparticle after
the neutralino. Figure \ref{m3bym2} demonstrates the correlation
between the GUT scale threshold corrections \eps\ and the ratio of
gaugino masses at the weak scale and the GUT scale.  Note, the
scalar masses are heavy and degenerate.  As a result, they do
not introduce differential running of the coupling constants. \\[5pt]

\begin{center}
\begin{figure}
\centering
\includegraphics[width=14.0cm]{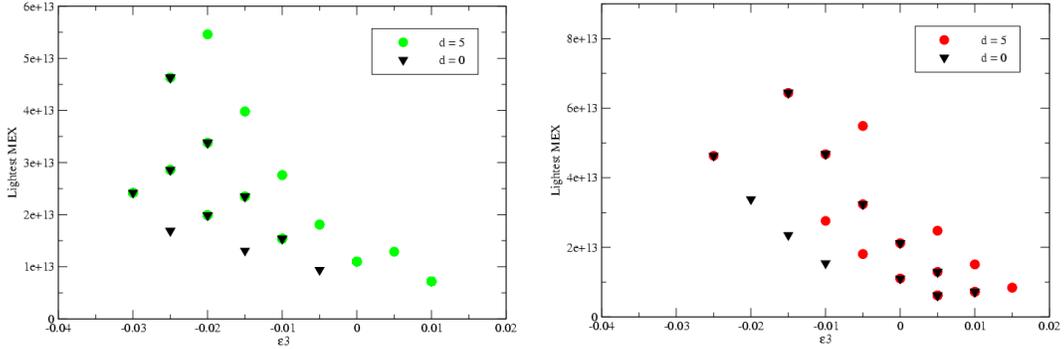}
\vspace{-50pt}
\caption{The plot shows the consistent points with varying \mgrav\ and d.
The plot on the left is for \mgrav\ = 4 TeV and
the one on the right is for \mgrav\ = 10 TeV. }
\label{higgs}
\end{figure}
\end{center}

\textit{SUSY breaking scale, gravitino mass, and D-Term}: The low
energy spectrum depends on the following parameters that are chosen
arbitrarily - the SUSY breaking scale $F$, the gravitino mass
\mgrav\,  and the $d$ parameter in the D-term.  Although the doublet
and the triplet exotics could couple to different SUSY breaking
fields, for simplicity we use a single SUSY breaking VEV, $F$. The
gravitino mass \mgrav, is the dominant contribution to the scalar
masses.  In order to obtain consistent solutions, we find \mgrav\
$\geq 2$ TeV, otherwise the scalars become non-degenerate at the GUT
scale and spoil unification through differential running. At the
same time, if \mgrav\ $>$ 10 TeV, the assumption that gauge
mediation is the dominant contribution to gaugino masses no longer
holds and the gravitino corrections to the gaugino masses must be
included. The D-term introduces a splitting between the sparticle
masses and the Higgs masses, since they carry different charges
under the $U_{X}(1)$. For the two cases of \mgrav\ = 4 TeV and 10
TeV, the graph \ref{higgs} shows the set of consistent points with
varying d = 0, 5.

\textit{MSSM Spectrum}:  The MSSM spectrum for the four particular
cases (Table \ref{gut_spectrum}) is given in Table
\ref{mssm_spectrum}. The \mgrav\ contribution makes the scalars very
heavy, with the third family being slightly lighter. The gauginos
receive the dominant contributions only from the gauge messengers
and are light in comparison with the scalars. The LSP is the
lightest neutralino, $\tilde{\chi}_{1}^{0}$, which is predominantly
``bino-like''. The gluino and chargino masses depend on the
threshold corrections at the GUT scale, as is seen for the four cases
given in Table \ref{mssm_spectrum}. Most of the points that were
found to be consistent with the low energy data have small values of
$\tan \beta$  $<$ 10. Finally, increasing the $d$ parameter gives a
handle on the possible values for $\tan \beta$. \\[5pt]

\textit{Collider prospects}: We have found light gauginos which can
be produced at the LHC or possibly the Tevatron. Since the gauginos
are lighter than the scalars, they will decay only through off-shell
squarks. Gluinos can decay via the process: $\tilde{g} \rightarrow q
\ \bar q \ \tilde{\chi}_{i}$. The produced $ \tilde{\chi}_{i}$ would
then undergo cascade decay until the final product is the LSP and
Standard Model leptons. This would give a striking signature of at
least 4 jets + missing $E_{T}$ \cite{Wells:2003tf}. The heavier
$\tilde{\chi}_{i}^{0}$ and $\tilde{\chi}_{i}^{\pm}$ could decay into
their lighter counterparts and leptons that could be the cleanest
signature at the LHC. The unique feature of the spectrum is the mass
difference between the heavier neutralinos and the LSP - about 150
GeV in one case and close to 500 GeV in the other. This could lead
to very high energy leptons and a lot of missing energy making this
a very favorable channel at the LHC. Once detected, this would give
useful information about the GUT scale threshold corrections.

\begin{table}
\centering
 \caption{\label{mssm_spectrum}The MSSM spectrum at the weak scale for the four different cases given in Table \ref{gut_spectrum}.}
 \vspace{5mm}
\begin{tabular}{|c|c|c|c|c|}
\hline
 \textbf{Observable} & \textbf{Case 1} & \textbf{Case 2} & \textbf{Case 3}  & \textbf{Case 4}\\ \hline
 tan $\beta$  & 7 & 4 & 4 & 6 \\
 mu   & -206.217 & -1932.930 &  937.044 & 958.984\\ \hline
$m_{h^{0}}$  & 119.311 & 117.384 & 117.791 & 117.384\\
$m_{H^{0}}$  & 4039.466 & 10327.840 &  10323.870 & 4129.50\\
$m_{A^{0}}$  & 4037.954& 10323.182 & 10319.566 & 4127.62\\
$m_{H^{+}}$  & 4039.133& 10323.750 &  10320.322 & 4128.69\\ \hline
$m_{\tilde{g}}$  & 708.987&455.343 &  369.786 & 712.66 \\  \hline
$m_{\tilde{\chi}_{1}^{0}}$  &52.867 & 60.336 &  13.608 & 24.671 \\
$m_{\tilde{\chi}_{2}^{0}}$  & 198.922& 548.051 & 97.766 & 629.15\\
$m_{\tilde{\chi}_{3}^{0}}$  & -221.679& 1947.429 & -947.105 & -962.34\\
$m_{\tilde{\chi}_{4}^{0}}$  & 360.330& -1949.000 & 952.206 & 974.12\\  \hline
$m_{\tilde{\chi}_{1}^{\pm}}$  & 199.225 & 548.196 & 98.136 & 629.87\\
$m_{\tilde{\chi}_{2}^{\pm}}$  & 351.551&1974.466 & 967.940 & 984.45\\   \hline
$m_{\tilde{d}_{L}} \simeq m_{\tilde{s}_{L}} $  & 4035.319& 10017.133 & 9891.84 & 4195.60\\
$m_{\tilde{u}_{L}} \simeq m_{\tilde{c}_{L}}$  &4034.710 & 10017.005 & 9891.67 & 4195.03\\
%$m_{\tilde{s}_{L}}$  &4035.305 & 10017.107\\
%$m_{\tilde{c}_{L}}$  & 4034.697& 10016.979 \\
$m_{\tilde{b}_{1}}$  & 3312.041& 8329.932 & 8180.58 & 3489.22 \\
$m_{\tilde{t}_{1}}$  & 2357.034& 6181.548 & 6034.69 & 2419.63\\ \hline
$m_{\tilde{e}_{L}} \simeq m_{\tilde{\mu}_{L}}$  & 4023.779& 10095.067 & 9973.61 & 4185.18 \\
$m_{\tilde{\nu}_{eL}} \simeq m_{\tilde{\nu}_{\mu L}}$  & 4022.711& 10094.474 & 9973.00 & 4184.15\\
%$m_{\tilde{\mu}_{L}}$  &4023.815 & 10095.059\\
%$m_{\tilde{\nu}_{\mu L}}$  & 4022.684& 10094.453\\
$m_{\tilde{\tau}_{L}}$  &3982.090 &  10068.311 & 9966.998 & 4053.27\\
$m_{\tilde{\nu}_{\tau L}}$  & 4014.664& 10088.009 & 9966.40 & 4178.30 \\ \hline
$m_{\tilde{d}_{R}} \simeq m_{\tilde{s}_{R}}$  &4011.295 & 10009.631 & 9919.58 & 4078.18 \\
$m_{\tilde{u}_{R}} \simeq m_{\tilde{c}_{R}}$  &4009.916 & 10005.369 & 9914.576 & 4076.57\\
%$m_{\tilde{s}_{R}}$  &4011.287 & 10009.625\\
%$m_{\tilde{c}_{R}}$  & 4009.897& 10005.322 \\
$m_{\tilde{b}_{2}}$  & 3994.851& 9997.358 & 9907.23 & 4065.98\\
$m_{\tilde{t}_{2}}$  &3314.600 & 8330.918 & 8181.58 & 3491.80\\ \hline
$m_{\tilde{e}_{R}}$  & 3998.343& 10081.311 & 9993.46 & 4065.36\\
$m_{\tilde{\mu}_{R}}$  & 3998.289& 10081.268  & 9993.42 & 4065.32\\
$m_{\tilde{\tau}_{R}}$  &4015.735 & 10088.622 & 9980.25 & 4179.34\\
\hline
\end{tabular}
\end{table}

\section{Summary}
We have calculated the spectrum of exotics as well the MSSM spectrum
starting from a heterotic string theory compactified on an
anisotropic orbifold. Allowing the exotics, that come in incomplete
GUT multiplets to obtain mass according to their quantum numbers
allows for more possible solutions to gauge coupling unification. We
find that we can build consistent MSSM spectra starting with such
theories with the exotics acting as messengers of SUSY breaking
through the gauge mediation mechanism. The gaugino masses in the low
energy spectrum depend on the threshold corrections at the GUT
scale. They are lighter than the scalars and are within the
kinematic reach of the LHC. The unique features of the spectrum are
light gluinos and also large mass differences between the lightest
and the next-to-lightest neutralinos and charginos which could lead
to interesting signatures at the LHC.
\\ [5pt]

\textbf{\large{Acknowledgments}}\\ [5pt] We would like to thank Ben
Dundee and Konstantin Bobkov for useful discussions. We also
received partial support for this work from DOE grant
DOE/ER/01545-891.

\newpage

\appendix
\section{\label{rge}Renormalization Group Equations}

\subsection{4D SUSY GUT Theory}
The grand unification (GUT) scale, \mgut\ is defined by SOFTSUSY
\cite{Allanach:2001kg} as the point where  \beqn
 \alpha_{1}(\mgut)  &=& \alpha_{GUT} \ = \ \alpha_{2}(\mgut) \nonumber \\
 \alpha_{3}(\mgut) &=& \alpha_{GUT} (1 + \eps) .
\eeqn Precision electroweak data requires, in the standard
supersymmetry breaking scenarios, with universal gaugino and scalar
masses at the GUT scale: \beq \eps = \frac{\alpha_{3} -
\alpha_{GUT}}{\alpha_{GUT}} \simeq -0.03 \label{4d} . \eeq  Thus the
1-loop renormalization group equations for SUSY GUT from the weak
scale to the GUT scale, with 2-loop threshold corrections near the
GUT scale are given (in the vicinity of the GUT scale) by: \beq
 \alpha_{i}^{-1} (\mu) = \alpha_{GUT}^{-1} + \frac{b_{i}}{2\pi} log
\frac{\mgut}{\mu} - \Delta \delta_{i3} \label{4drg} \eeq where i =
3,2,1, represent $SU(3)_{C}$, $SU(2)_{L}$, and $U(1)_{Y}$
respectively. The $b_{i}s$ are the $\beta$-function coefficients,
$b_{i} = (-3, 1, \frac{33}{5})$ for the MSSM and $\delta_{i3}$ is
the threshold corrections to $\alpha_{3}$, \beq \Delta =
\alpha_{GUT}^{-1} \frac{\eps}{(1 + \eps)}. \eeq  It should be noted
that when we work in the non-standard scenarios such as models with
non-universal gaugino masses at the GUT scale, the value of \eps\
need not be -3 \%.

\subsection{Orbifold GUT Theory}
The RGEs for the orbifolded model can be arrived at by taking into
account all the particles in the spectrum. The highest scale in this
theory is the string scale,  \ms, above which there is one unified
grand unified coupling constant, $\alpha_{string}$. In the heterotic
framework, the unified coupling constant is related to Newton's
constant, $G_{N}$, by the relation: \beq G_{N} = \frac{1}{8} \
\alpha_{\text{string}} \ \alpha' \eeq where, $\alpha' =
\frac{1}{\ms^{2}}$, and from the observed value of $G_{N} =
\frac{1}{M_{PL}^{2}}$; $M_{PL} \sim 1.2 \times 10^{19}$ GeV. Thus we
have, \beq \alpha_{string} = 8 \frac{\ms^{2}}{M_{PL}^{2}}
\label{string} \eeq The three gauge couplings renormalize
independently below the string scale. New states enter the theory at
the intermediate scales:
\begin{itemize}
 \item \mext\ - Mass scale of the triplet exotics.
 \item \mexd\ - Mass scale of the doublet exotics.
 \item \mc\ - The compactification scale of the extra-dimensions.
\end{itemize}

 \begin{figure}[ht]
 \centering
 \includegraphics[width=0.85\textwidth]{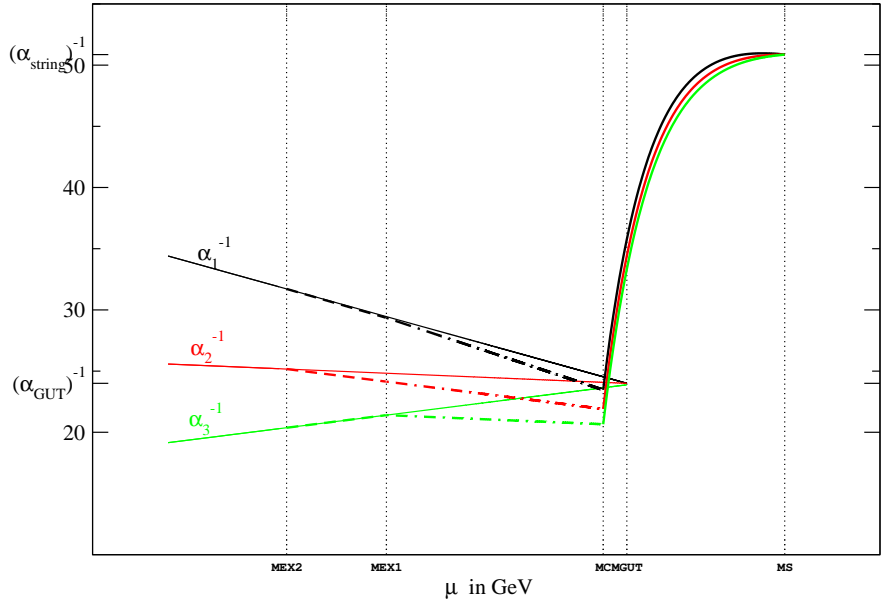}
 \caption{The figure shows the evolution of couplings for one particular model
$\vec{n}$ = (4,2,(2,1)). The 4D GUT scale is \mgut\ = $2 \times 10^{16}$ GeV.
The doublet exotics enter at \mexd\ = $1.29 \times 10^{13}$ GeV; the triplets
enter at \mext\ = $1.115 \times 10^{14}$ GeV. At \mc\ = $1.2 \times 10^{16}$
GeV, the Kaluza-Klein states of the MSSM particles enter and couplings run with a
power law to unify at the string scale, \ms\ =$6.06 \times 10^{17}$ GeV. }
 \label{running}
\end{figure}

In general each of the exotics could obtain different mass. We
generalize from the previous work \cite{Dundee:2008ts}, where it was
assumed that all the exotics obtain mass at the same scale. This
generalization is motivated by the fact that the exotics do not come
in complete GUT multiplets. Hence the possibility of a splitting
between the doublets and triplets is well motivated. The running of
the couplings begin to alter at these scales, as shown in Fig.
\ref{running}. Taking into account all particles in the spectrum
below the string scale: MSSM particles, exotic states and the
Kaluza-Klein modes from the compactification, the renormalization
group equation for the orbifold GUT theory becomes: \beqn
 \alpha_{i}^{-1} (\mu) = \alpha_{string}^{-1} &+& \frac{b_{i}^{MSSM}}{2\pi} log
\frac{M_{s}}{\mu}+\frac{b^{EX3}_{i}}{2\pi} log
\frac{M_{s}}{M_{EX3}}+ \frac{b^{EX2}_{i}}{2\pi} log
\frac{M_{s}}{M_{EX2}}\nonumber \\ &-& \frac{1}{4
\pi}(b_{i}^{++}+b_{i}^{--})log \frac{M_{s}}{M_{c}} +
\frac{b^{G}}{2\pi}(\frac{M_{s}}{M_{c}}-1). \label{orbrge} \eeqn The
first term in the above equation is the tree level boundary
condition from the heterotic string theory as given in Eq.
(\ref{string}). The next three terms include contributions to the
$\beta$-function coefficients from the from the MSSM and exotic
brane states and the zero-KK modes of the states living in the bulk.
\begin{align}
b_{i}^{X} = b_{i}^{X, ++} + b_{i}^{X, \text{brane}} \qquad  \text{where}\ X =
MSSM,\ EX3,\ EX2.
\end{align}
For the MSSM, $b_{i} = (3, -1, 33/5)$.  The infinite sum over the KK
modes of the MSSM can be evaluated to give the last term in Eq.
(\ref{orbrge}) \cite{Dienes:1998vg}. States that are not localized
on the ``branes'' are free to propagate in the higher-dimensional
``bulk''. In this work, we have considered that only the minimum
amount of matter lives in the bulk - the MSSM third family $\bar{b}$
and L. In principle we could allow two more pairs of chiral
multiplets, $\mathbf{6} + \mathbf{6^{c}}$ in the bulk that would
correspond to ``bulk'' exotics. The contribution from the massive KK
modes of the exotics that live in the bulk would also include an
infinite sum over the KK modes. The case of bulk exotics was
analyzed in \cite{Dundee:2008ts} and the spectrum of exotics was
calculated. For the purposes of the current work, we shall assume
that the exotics live only on the brane. The current treatment can
be extended similarly for the bulk exotics case also, without any
significant change in the low energy phenomenology.

The brane-localized exotics' $\beta$-function coefficients depend on
the exotic matter content of the theory, see Table
\ref{tab:all_exotics}. The exotic matter content is defined in terms
of the parameters $n_{i}$ with: \beqn
 n_{3}\times
\left[(\textbf{3},1)_{1/3,*}+(\overline{\textbf{3}},1)_{-1/3,*}\right]
+ n_{2} \times
\left[(1,\textbf{2})_{0,*}+(1,\textbf{2})_{0,*}\right] + \nonumber
\\ n_1 \times \left[(1,1)_{1,*}+(1,1)_{-1,*}\right] \eeqn The
coefficients \begin{align*}
 b^{EX3} = (n_{3}, 0, \frac{n_{3} + 3 n_{1}}{10}) \qquad b^{EX2} = (0, n_{2},
\frac{3 n'_{1}}{10})
\end{align*}(given in table \ref{betavalues}) can be calculated for
each different value of the parameters $\overrightarrow{n} = (n_{3},
n_{2}, (n_{1}, n'_{1}))$.  In the above expression,  $n_{1}$ of the
exotics with just U(1) hypercharge obtain the same mass as the
triplets and $n'_{1}$ of them obtain the same mass as the doublets.
We check for gauge coupling unification by comparing the following
quantities (i) $1/\alpha_{3}-1/\alpha_{2}$,  (ii) $
1/\alpha_{2}-1/\alpha_{1}$, (iii) $ \alpha_{3} $ from the 4D SUSY
GUT theory and the orbifold GUT theory at the scale $\mu$. We scan
over all possible, $n_i$, and check for gauge coupling unification.

\begin{table}[ht]
 \centering
 \caption{\label{exotic_spectrum}Exotic matter content in Models 1A/B and 2 from
\cite{minilandscape}. Listed are the quantum numbers of the states
under the MSSM and hidden sector gauge groups, with the hypercharge
denoted in the subscript. The brane localized exotic matter in Model
1 is a subset of that in Model 2.}
 \vspace{5mm}
 \label{tab:all_exotics} \begin{footnotesize}
 \begin{tabular}{|c|c|c|c|c|}
 \hline
 Model&Hidden Sector&&Exotic Matter Irrep&Name\\
 \hline
 \hline
1 A/B   &$SU(4) \times SU(2)$    &brane  & $2 \times
\left[(\textbf{3},1;1,1)_{1/3,2/3}+(\overline{\textbf{3}},1;1,1)_{-1/3,-2/3}
\right]$    & $v+\bar{v}$\\
    &           &exotics& $4 \times
\left[(1,\textbf{2};1,1)_{0,*}+(1,\textbf{2};1,1)_{0,*}\right]$             & $m
+ m$\\
    &           &   & $1 \times
\left[(1,\textbf{2};1,\textbf{2})_{0,0}+(1,\textbf{2};1,\textbf{2})_{0,0}\right]
$           & $y + y$\\
    &           &   & $2 \times
\left[(1,1;\mathbf{4},1)_{1,1}+(1,1;\overline{\mathbf{4}},1)_{-1,-1}\right]$
& $f^+ + \bar{f}^-$\\
    &           &   & $14 \times \left[(1,1;1,1)_{1,*}+(1,1;1,1)_{-1,*}\right]$
               & $s^+ + s^-$\\
\hline
    &           &bulk   &$6 \times
\left[(\textbf{3},1;1,1)_{-2/3,-2/3}+(\overline{\textbf{3}},1;1,1)_{2/3,2/3}
\right] $    &$\delta + \bar{\delta}$\\
    &           &exotics&$1 \times
\left[(\textbf{3},1;1,1)_{-2/3,-1/3}+(\overline{\textbf{3}},1;1,1)_{2/3,1/3}
\right] $    &$d + \bar{d}$\\
    &           &   &$1
\times\left[(1,\textbf{2};1,1)_{-1,-1}+(1,\textbf{2};1,1)_{1,1}\right]$
   &$\ell + \bar{\ell}$\\
%\hline
\hline
2   & $SO(8)\times SU(2)$     &brane  & $4 \times
\left[(\textbf{3},1;1,1)_{1/3,*}+(\overline{\textbf{3}},1;1,1)_{-1/3,*}\right] $
   & $v+\bar{v}$\\
    &           &exotics& $2 \times
\left[(1,\textbf{2};1,1)_{0,*}+(1,\textbf{2};1,1)_{0,*}\right]$             & $m
+ m$\\
    &           &   & $1 \times
\left[(1,\textbf{2};1,\textbf{2})_{0,0}+(1,\textbf{2};1,\textbf{2})_{0,0}\right]
$           & $y + y$\\
    &           &   & $2 \times
\left[(1,1;1,\textbf{2})_{1,1}+(1,1;1,\textbf{2})_{-1,-1}\right]$
& $x^+ + x^-$\\
    &           &   & $20 \times \left[(1,1;1,1)_{1,*}+(1,1;1,1)_{-1,*}\right]$
               & $s^+ + s^-$\\
\hline
    &           &bulk   & $3 \times
\left[(\textbf{3},1;1,1)_{-2/3-2/3}+(\overline{\textbf{3}},1;1,1)_{2/3,2/3}
\right]$ &$\delta + \bar{\delta}$\\
    &           &exotics&$1 \times
\left[(\textbf{3},1;1,1)_{-2/3,2/3}+(\overline{\textbf{3}},1;1,1)_{2/3,-2/3}
\right]$ &$d + \bar{d}$\\
    &           &   &$1
\times\left[(1,\textbf{2};1,1)_{-1,-1}+(1,\textbf{2};1,1)_{1,1}\right]$
   &$\ell + \bar{\ell}$\\
    &           &   &$3
\times\left[(1,\textbf{2};1,1)_{-1,0}+(1,\textbf{2};1,1)_{1,0}\right]$
   &$\phi + \bar{\phi}$\\
 \hline
 \end{tabular}
 \end{footnotesize}
\end{table}

\begin{table}
 \begin{center}
 \begin{tabular}{|c|c|c|c|}
 \hline
  irrep & $b_{3}$ & $b_{2}$ & $b_{Y}$ \\ \hline \hline
  $\left[(\textbf{3},1)_{1/3,*}+(\overline{\textbf{3}},1)_{-1/3,*}\right]$ & 1 &
0 & 1/10 \\
  $\left[(1,\textbf{2})_{0,*}+(1,\textbf{2})_{0,*}\right] $ & 0& 1 & 0 \\
 $ \left[(1,1)_{1,*}+(1,1)_{-1,*}\right] $ & 0 & 0 & 3/10 \\
 \hline
 \end{tabular}
 \caption{Values of $\beta$-function coefficients for the brane-localized
exotics.}
 \label{betavalues}
  \end{center}
\end{table}

\subsection{Spectrum of Exotics and Constraints}
We look for models with gauge coupling unification allowing the
triplets and the doublets to obtain mass at different scales: \mext,
and \mexd. The relevant quantities in the RGEs are given by:

\begin{align}
 \text{MSSM $\beta$-function coefficients} \qquad&  b^{MSSM} = (-3, 1,
\frac{33}{5}) &\nonumber \\
 \text{MSSM bulk states $\beta$-function coefficients} \qquad&  b^{++} = (-7,
-3, \frac{13}{5})& \nonumber \\
 \qquad& b^{--} = (5, 1, \frac{1}{5}) & \nonumber \\
 b^{G} \equiv \sum_{P=\pm, P'=\pm} b_{i}^{MSSM, PP'}; \qquad& b^{G} = -4 &
\nonumber \\
 \text{Planck Scale} \qquad&  M_{PL} \  =\ 1.22 \times 10^{19}\  \text{GeV}  &
\nonumber
\end{align}

\mc\ is allowed to vary between a minimum of $7 \times 10^{15}$ GeV,
the approximate bound determined in \cite{Dundee:2008ts} and higher.
We find the maximum number of solutions for \mc\ = $1.2 \times
10^{16}$ GeV. All further results shown will be from the data set
with this value of \mc. This also seems to be a good choice of \mc\
since it is only an order of magnitude smaller than the string
scale. Since we are deviating from the standard scenario of
universal gaugino masses at the GUT scale, we need to consider a
larger range of \eps. In fact, we assume that the GUT threshold
corrections, in order to agree with the precision electroweak data,
can be anywhere between \eps\ $\sim$ -4\% and +2\%. In comparison
with the earlier analysis, \cite{Dundee:2008ts}, we find a larger
number of cases that satisfy gauge coupling unification. Introducing
the splitting of the doublets and triplets gives more freedom in
fixing the compactification scale, and hence a larger number of
solutions fall within the proton lifetime bound (See eq.
\ref{proton}).

\subsubsection{Proton Decay}
The bound on the proton lifetime is an important constraint on the
solutions, since the models discussed here are SU(6) GUTS in 5D. At
the scale, $M_{X} \sim  M_{C}$, the grand unified gauge group is
broken and the GUT gauge \textbf{X} boson gets mass at this scale.
It can mediate proton decay via dimension 6 operators. In a 4D
effective theory formalism, the decay rate $\Gamma (p \rightarrow
e^{+} + \pi^{0})$ can be calculated in terms of $g_{GUT}$, \ms, \mc,
and $M_{X}$. In Appendix B of Ref. \cite{Dundee:2008ts} they obtain
\beq \tau(p \rightarrow e^{+}\pi^{0}) = 5.21 \times 10^{40} \left(
\frac{\mc}{\ms} \right) ^{4} \text{years}. \label{proton} \eeq

When calculating the spectrum of exotics satisfying gauge coupling
unification, we only keep cases consistent with the experimental
bounds on the proton lifetime. The strongest experimental bound
comes from Super-Kamiokande in Japan that searches for  $ p
\rightarrow \ e^{+} + \pi^{0}$ signatures. The current bound is
\cite{ichep10} \beq \tau(p \rightarrow e^{+}\pi^{0}) > 1 \times
10^{34} \text{years}. \eeq

\subsubsection{\label{gcoupling} String Coupling Constant}
We have assumed that the $E_{8} \times E_{8}$  model considered in
this paper is in the weakly coupled regime. In \cite{Dundee:2008tr}
a simple formula relating the string coupling constant,
$g_{string}$, to other relevant quantities of the model was obtained
\beq g_{string}^{2} = \alpha_{GUT} \frac{\ms}{\mc}. \eeq  To be in
the perturbative regime, we need $g_{string} \leq$ 1. In the
heterotic framework, we also have a relation between
$\alpha_{GUT}(\ms) \equiv \alpha_{string}$, i.e. the unified
coupling constant at the string scale, given in Eq. \ref{string}:
\beq \alpha_{GUT}^{-1} = \frac{1}{8} \frac{M_{PL}^{2}}{\ms^{2}}.
\eeq This condition then imposes an additional constraint on the
possible consistent solutions. The value of $g_{string}$ 
for the four cases discussed earlier can be found in Table. \ref{gut_spectrum}.

\section{\label{2lc}2-loop corrections to soft masses}
The 2-loop renormalization group equations for the MSSM parameters
have been studied extensively in literature \cite{2-loop}. The RGEs
for the couplings and gaugino masses at 2-loop are given by: \beqn
\frac{d M_{i}}{dt} &=& \frac{2 g_{i}^{2}}{16 \pi^{2}} b_{i}^{(1)}
M_{i} + \frac{2 g_{i}^{2}}{(16 \pi^{2})^{2}} \sum_{j=1}^{3}
b_{ij}^{(2)} g_{j}^{2}
(M_{i} + M_{j} ) \\
\frac{d g_{i}}{dt} &=& \frac{g_{i}^{3}}{16 \pi^{2}} b_{i}^{(1)} +
\frac{g_{i}^{3}}{(16 \pi^{2})^{2}} \sum_{j=1}^{3} b_{ij}^{(2)}
g_{j}^{2} \eeqn where, $\frac{g_{i}^{2}}{4 \pi} = \alpha_{i}$. The
1-loop and 2-loop $\beta$- function coefficients for the MSSM are:
\begin{align}
 b_{i}^{(1)} = (-33/5, -1, 3) \qquad b_{ij}^{(2)} = \left( \begin{array}{ccc}
199/100 & 27/20 & 22/5 \\ 9/20 & 25/4 & 6 \\ 11/20 & 9/4 & 7/2 \end{array}
\right)
\end{align}
At 1-loop the gaugino masses and the couplings obey the relation:
\beq
\frac{M_{i}}{\alpha_{i}} = constant
\label{1loop}
\eeq
At 2-loops this equation gets small corrections. Solving the above two equations
at 2-loops, we find:
\beq
\frac{M_{i}}{\alpha_{i}} = \sum_{j=1}^{3} b_{ij}^{(2)} \alpha_{j} M_{j}
\eeq
Approximating the right-hand side by the 1-loop result, and using
the values of the $\beta$-function coefficients, we find that the
2-loop corrections to Eq. (\ref{1loop}) are less than 1\%
\cite{Yamada:1994si}.


\newpage



\begin{thebibliography}{10}
\bibitem{Georgi:1974sy}
  H.~Georgi, S.~L.~Glashow,
  ``Unity of All Elementary Particle Forces,''
  Phys.\ Rev.\ Lett.\  {\bf 32}, 438-441 (1974).

\bibitem{Georgi:1974my}
  H.~Georgi,
  ``The State of the Art - Gauge Theories. (Talk),''
  AIP Conf.\ Proc.\  {\bf 23}, 575-582 (1975).

 H.~Fritzsch, P.~Minkowski,
  ``Unified Interactions of Leptons and Hadrons,''
  Annals Phys.\  {\bf 93}, 193-266 (1975).

\bibitem{Dimopoulos:1981yj}
S.~Dimopoulos, S.~Raby, and F.~Wilczek, ``Supersymmetry and the scale of
  unification,'' {\em Phys. Rev.} {\bf D24} (1981)
1681--1683.
%%CITATION = PHRVA,D24,1681;%%.

S.~Dimopoulos and H.~Georgi, ``{Softly Broken Supersymmetry and SU(5)},'' {\em
  Nucl. Phys.} {\bf B193} (1981)
150.
%%CITATION = NUPHA,B193,150;%%.

L.~E. Ibanez and G.~G. Ross, ``{Low-Energy Predictions in Supersymmetric Grand
  Unified Theories},'' {\em Phys. Lett.} {\bf B105} (1981)
439.
%%CITATION = PHLTA,B105,439;%%.

N.~Sakai, ``{Naturalness in Supersymmetric Guts},'' {\em Zeit. Phys.} {\bf C11}
  (1981)
153.
%%CITATION = ZEPYA,C11,153;%%.

M.~B. Einhorn and D.~R.~T. Jones, ``{The Weak Mixing Angle and Unification Mass
  in Supersymmetric SU(5)},'' {\em Nucl. Phys.} {\bf B196} (1982)
475.
%%CITATION = NUPHA,B196,475;%%.

W.~J. Marciano and G.~Senjanovi\'c, ``{Predictions of Supersymmetric Grand
  Unified Theories},'' {\em Phys. Rev.} {\bf D25} (1982)
3092.
%%CITATION = PHRVA,D25,3092;%%.

\bibitem{Amsler:2008zzb}
  C.~Amsler {\it et al.} [Particle Data Group Collaboration] ,
  %``Review of Particle Physics,''
  Phys.\ Lett.\  {\bf B667}, 1 (2008).

\bibitem{kazakov} 

D.I. Kazakov, Lectures given at the European School on High Energy Physics, Aug.-Sept. 2000, Caramulo,
Portugal [hep-ph/0012288v2].

%\cite{Lucas:1995ic}
  V.~Lucas and S.~Raby,
  ``GUT Scale Threshold Corrections in a Complete Supersymmetric SO(10) Model:
  $\alpha_s(M_Z)$ vs. Proton Lifetime,''
{\it Phys. Rev.} {\bf D54}, 2261 (1996)
  [arXiv:hep-ph/9601303];
  %\cite{Blazek:1996yv}
%\bibitem{Blazek:1996yv}
 
 T.~Blazek, M.~Carena, S.~Raby and C.~E.~M.~Wagner,
``A global chi**2 analysis of electroweak data in SO(10) SUSY GUTs,''
 {\it Phys. Rev.}  {\bf D56}, 6919 (1997)
  [arXiv:hep-ph/9611217];
%\cite{Altarelli:2000fu}
%\bibitem{Altarelli:2000fu}

  G.~Altarelli, F.~Feruglio and I.~Masina,
``From minimal to realistic supersymmetric SU(5) grand unification,''
{\it  JHEP} {\bf 0011}, 040 (2000)
  [arXiv:hep-ph/0007254].

 R. Derm\' \i \v sek, A. Mafi and S. Raby, {\it Phys. Rev.} {\bf D63}, 035001 (2001);

K.S. Babu, J.C. Pati and F. Wilczek, {\it Nucl. Phys.} {\bf B566}, 33 (2000).

%\cite{Alciati:2005ur}

  M.~L.~Alciati, F.~Feruglio, Y.~Lin and A.~Varagnolo,
``Proton lifetime from SU(5) unification in extra dimensions,''
{\it  JHEP} {\bf 0503}, 054 (2005)
  [arXiv:hep-ph/0501086].


\bibitem{Dundee:2008ts}
  B.~Dundee, S.~Raby, A.~Wingerter,
  %``Reconciling Grand Unification with Strings by Anisotropic
Compactifications,''
  Phys.\ Rev.\  {\bf D78}, 066006 (2008).
  [arXiv:0805.4186 [hep-th]].
\bibitem{Raby:2009sf}
  S.~Raby, M.~Ratz and K.~Schmidt-Hoberg,
  %``Precision gauge unification in the MSSM,''
  Phys.\ Lett.\  B {\bf 687}, 342 (2010)
  [arXiv:0911.4249 [hep-ph]].
  %%CITATION = PHLTA,B687,342;%%
\bibitem{ichep10}
  Makoto Miura [Super-Kamiokande Collaboration],
  ''Search for Nucleon Decays in Super-Kamiokande``
  %``Search for Proton Decay via p -> e^+ pi^0 and p -> mu^+ pi^0 in a Large
  %Water Cherenkov Detector,''
  ICHEP 2010.
 % Phys.\ Rev.\ Lett.\  {\bf 102}, 141801 (2009)
  %[arXiv:0903.0676 [hep-ex]].
  %%CITATION = PRLTA,102,141801;%%
\bibitem{Giudice:1998bp}
  G.~F.~Giudice and R.~Rattazzi,
  %``Theories with gauge-mediated supersymmetry breaking,''
  Phys.\ Rept.\  {\bf 322}, 419 (1999)
  [arXiv:hep-ph/9801271].
  %%CITATION = PRPLC,322,419;%%
\bibitem{Martin:1997ns}
  S.~P.~Martin,
  \textit{``A Supersymmetry Primer''},
  arXiv:hep-ph/9709356.
  %%CITATION = HEP-PH/9709356;%%
\bibitem{Allanach:2001kg}
  B.~C.~Allanach,
  %``SOFTSUSY: A C++ program for calculating supersymmetric spectra,''
  Comput.\ Phys.\ Commun.\  {\bf 143}, 305 (2002)
  [arXiv:hep-ph/0104145].
  %%CITATION = CPHCB,143,305;%%
\bibitem{gauge-higgs}
W.~Buchmuller, K.~Hamaguchi, O.~Lebedev and M.~Ratz,
  %``Supersymmetric standard model from the heterotic string,''
  Phys.\ Rev.\ Lett.\  {\bf 96}, 121602 (2006)
  [arXiv:hep-ph/0511035].

 W.~Buchmuller, K.~Hamaguchi, O.~Lebedev and M.~Ratz,
  %``Supersymmetric standard model from the heterotic string. II,''
  Nucl.\ Phys.\  B {\bf 785}, 149 (2007)
  [arXiv:hep-th/0606187].

 O.~Lebedev, H.~P.~Nilles, S.~Raby, S.~Ramos-Sanchez, M.~Ratz,
P.~K.~S.~Vaudrevange and A.~Wingerter,
  %``Low Energy Supersymmetry from the Heterotic Landscape,''
  Phys.\ Rev.\ Lett.\  {\bf 98}, 181602 (2007)
  [arXiv:hep-th/0611203].

\bibitem{minilandscape}
  O.~Lebedev, H.~P.~Nilles, S.~Raby, S.~Ramos-Sanchez, M.~Ratz,
P.~K.~S.~Vaudrevange and A.~Wingerter,
  %``A mini-landscape of exact MSSM spectra in heterotic orbifolds,''
  Phys.\ Lett.\  B {\bf 645}, 88 (2007)
  [arXiv:hep-th/0611095].

\bibitem{Dienes:1998vg}
  K.~R.~Dienes, E.~Dudas and T.~Gherghetta,
  %``Grand unification at intermediate mass scales through extra dimensions,''
  Nucl.\ Phys.\  B {\bf 537}, 47 (1999)
  [arXiv:hep-ph/9806292].
  %%CITATION = NUPHA,B537,47;%%

  \bibitem{Raby:1998xr}
  S.~Raby and K.~Tobe,
  %``The phenomenology of SUSY models with a gluino LSP,''
  Nucl.\ Phys.\  B {\bf 539}, 3 (1999)
  [arXiv:hep-ph/9807281].
  %%CITATION = NUPHA,B539,3;%%

\bibitem{Wells:2003tf}
  J.~D.~Wells,
  %``Implications of supersymmetry breaking with a little hierarchy between
  %gauginos and scalars,''
  arXiv:hep-ph/0306127.
  %%CITATION = HEP-PH/0306127;%%

  %\cite{Dundee:2008tr}
\bibitem{Dundee:2008tr}
  B.~Dundee and S.~Raby,
  %``On the string coupling in a class of stringy orbifold GUTs,''
  arXiv:0808.0992 [hep-th].
  %%CITATION = ARXIV:0808.0992;%%

  %\cite{Dine:1981gu}
\bibitem{gaugemediation}

  M.~Dine and W.~Fischler,
  %``A Phenomenological Model Of Particle Physics Based On Supersymmetry,''
  Phys.\ Lett.\  B {\bf 110}, 227 (1982).
  %%CITATION = PHLTA,B110,227;%%

    C.~R.~Nappi and B.~A.~Ovrut,
  %``Supersymmetric Extension Of The SU(3) X SU(2) X U(1) Model,''
  Phys.\ Lett.\  B {\bf 113}, 175 (1982).
  %%CITATION = PHLTA,B113,175;%%

    L.~Alvarez-Gaume, M.~Claudson and M.~B.~Wise,
  %``Low-Energy Supersymmetry,''
  Nucl.\ Phys.\  B {\bf 207}, 96 (1982).
  %%CITATION = NUPHA,B207,96;%%

\bibitem{2-loop}

  S.~P.~Martin and M.~T.~Vaughn,
  %``Two Loop Renormalization Group Equations For Soft Supersymmetry Breaking
  %Couplings,''
  Phys.\ Rev.\  D {\bf 50}, 2282 (1994)
  [Erratum-ibid.\  D {\bf 78}, 039903 (2008)]
  [arXiv:hep-ph/9311340].
  %%CITATION = PHRVA,D50,2282;%%

  Y.~Yamada,
  %``Two loop renormalization group equations for soft SUSY breaking scalar
interactions: Supergraph method,''
  Phys.\ Rev.\  {\bf D50}, 3537-3545 (1994).
  [hep-ph/9401241].

  I.~Jack, D.~R.~T.~Jones,
  %``Soft supersymmetry breaking and finiteness,''
  Phys.\ Lett.\  {\bf B333}, 372-379 (1994).
  [hep-ph/9405233].

\bibitem{Yamada:1994si}
  Y.~Yamada,
  %``Two loop renormalization of gaugino masses,''
  In *Tsukuba 1993, Proceedings, Particle physics* 352-358.

\bibitem{Raby:1998bg}
  S.~Raby and K.~Tobe,
  %``Dynamical SUSY breaking with a hybrid messenger sector,''
  Phys.\ Lett.\  B {\bf 437}, 337 (1998)
  [arXiv:hep-ph/9805317].
  %%CITATION = PHLTA,B437,337;%%



\end{thebibliography}
\end{document}